%% file: main.tex
\documentclass[fleqn,usenatbib]{mnras}

\usepackage{newtxtext,newtxmath}
\usepackage[T1]{fontenc}
\usepackage{ae,aecompl}

\usepackage{graphicx}
\usepackage{amsmath}
\usepackage{amssymb}
\usepackage{gensymb}
\usepackage{threeparttable}
\usepackage{booktabs,siunitx}
\usepackage{tabularx}
\usepackage{dcolumn}
\newcolumntype{d}[1]{D{.}{.}{#1}}
\newcommand\dcolhead[1]{\multicolumn{1}{c}{#1}}

\newcommand\changes[1]{#1}%\textcolor{red}{\textbf{#1}}}

\title[Radio observations of NS merger outflows]{Constraining properties of neutron star merger outflows with radio observations}

\author[D. Dobie et al.]{Dougal Dobie,$^{1,2}$\thanks{E-mail: ddob1600@uni.sydney.edu.au (DD)}
David L. Kaplan,$^{3,4}$
Kenta Hotokezaka,$^{5,6}$
Tara Murphy,$^{1,7}$
\newauthor
Adam Deller,$^{8,7}$
Gregg Hallinan,$^{9}$
Samaya Nissanke,$^{10,11}$
\\
$^{1}$Sydney Institute for Astronomy, School of Physics, University of Sydney, NSW 2006, Australia\\
$^{2}$ATNF, CSIRO Astronomy and Space Science, PO Box 76, Epping, NSW 1710, Australia\\
$^{3}$Department of Physics, University of Wisconsin-Milwaukee, Milwaukee, WI 53201, USA\\
$^{4}$Department of Astronomy, Oskar Klein Centre, Stockholm University, AlbaNova University Centre, SE-106 91 Stockholm, Sweden\\
$^{5}$Department of Astrophysical Sciences, Princeton University, 4 Ivy Lane, Princeton, NJ 08544, USA\\
$^{6}$Research Center for the Early Universe, Graduate School of Science, University of Tokyo, Bunkyo-ku, Tokyo 113-0033, Japan\\
$^{7}$ARC Centre of Excellence for Gravitational Wave Discovery (OzGrav), Hawthorn, Victoria, Australia\\
$^{8}$Centre for Astrophysics and Supercomputing, Swinburne University of Technology, Hawthorn, Victoria,
Australia\\
$^{9}$Cahill Center for Astronomy \& Astrophysics, Caltech, Pasadena CA, USA\\
$^{10}$GRAPPA, Anton Pannekoek Institute for Astronomy and Institute of High-Energy Physics,\\University of Amsterdam, Science Park 904, 1098 XH Amsterdam, The Netherlands\\
$^{11}$Nikhef, Science Park 105, 1098 XG Amsterdam, The Netherlands\\
}

\date{Accepted XXX. Received YYY; in original form ZZZ}

\pubyear{2020}

\begin{document}
\label{firstpage}
\pagerange{\pageref{firstpage}--\pageref{lastpage}}
\maketitle

\begin{abstract}
The jet opening angle and inclination of GW170817 -- the first detected binary neutron star merger -- were vital to understand its energetics, relation to short gamma-ray bursts, and refinement of the standard siren-based determination of the Hubble constant, $H_0$. These basic quantities were determined through a combination of the radio lightcurve and Very Long Baseline Interferometry (VLBI) measurements of proper motion. In this paper we discuss and quantify the prospects for the use of radio VLBI observations and observations of scintillation-induced variability to measure the source size and proper motion of merger afterglows, and thereby infer properties of the merger including inclination angle, opening angle and energetics. We show that these techniques are complementary as they probe different parts of the circum-merger density/inclination angle parameter space and different periods of the temporal evolution of the afterglow. We also find that while VLBI observations will be limited to the very closest events it will be possible to detect scintillation for a large fraction of events beyond the range of current gravitational wave detectors. Scintillation will also be detectable with next generation telescopes such as the Square Kilometre Array, 2000 antenna Deep Synoptic Array and the next generation Very Large Array, for a large fraction of events detected with third generation gravitational wave detectors. Finally, we discuss prospects for the measurement of the $H_0$ with VLBI observations of neutron star mergers and compare this technique to other standard siren methods.
\end{abstract}

\input{lit_table}
\section{Introduction}
The first detection of gravitational waves and electromagnetic radiation from a neutron star merger \citep[GW170817;][]{2017PhRvL.119p1101A,2017ApJ...848L..12A,2017ApJ...848L..13A} has given insight into high energy astrophysics, nuclear physics and cosmology. Observations of the radio lightcurve of GW170817 were able to place constraints on merger parameters including the isotropic equivalent energy of the merger, the density of the surrounding environment, and the jet opening angle \citep{2017Sci...358.1579H,2017ApJ...850L..21K,2018Natur.554..207M,2018ApJ...858L..15D,2018ApJ...863L..18A,2018ApJ...856L..18M,2018ApJ...868L..11M,2018ApJ...867...57R,2018MNRAS.478L..18T,2019MNRAS.489.1919T,2018ApJ...869...55W,2019ApJ...886L..17H,2019MNRAS.490.2822Z}. However, observations of the radio lightcurve alone were unable to distinguish between two competing models for the geometry of the outflow \citep{2018MNRAS.478..407N}\changes{, although the steep decline of the lightcurve did slightly favour the presence of a jet \citep{2018MNRAS.478L..18T,2018MNRAS.481.2581L}}. This tension was not resolved until Very Long Baseline Interferometry (VLBI) observations of the afterglow detected superluminal motion, suggesting that the late-time radio emission in GW170817 was jet-dominated \citep{2018Natur.561..355M,2019Sci...363..968G}.

The observation of superluminal motion has also placed tighter constraints on the inclination angle of the merger. In turn, this helped break the distance-inclination degeneracy \citep{1993PhRvD..47.2198F,2010ApJ...725..496N} in the gravitational wave observations, which contributed to most of the error budget in the initial standard siren measurement of the Hubble constant, $H_0$, using GW170817 \citep{2017Natur.551...85A}. This allowed for a measurement of the $H_0$ with a precision of 7\% \citep{2019NatAs...3..940H} compared to 17\% using the gravitational wave data alone \citep{2017Natur.551...85A}. VLBI observations of $\sim 15$ similarly favourably oriented events with comparable signal-to-noise as GW170817 (combined with improvements in jet modelling and calibration of gravitational wave detectors) will allow for a measurement of $H_0$ with sufficient precision (<2\%) and accuracy \citep{2019arXiv190908627M} to potentially resolve the discrepancy \citep{2019NatAs...3..891V} between current estimates from cosmic microwave background power spectrum measurements \citep{2018arXiv180706209P} and distance ladder observations \citep{2018ApJ...861..126R,2019ApJ...876...85R,2019ApJ...886L..27R}. In comparison, it will take tens to hundreds of events to achieve a similar level of precision with gravitational wave observations alone \citep{1986Natur.323..310S,2012PhRvD..86d3011D,2012PhRvL.108i1101M,2012PhRvD..85b3535T,2019ApJ...883L..42F} and gravitational wave observations with independent redshift measurements \citep{2006PhRvD..74f3006D,2010ApJ...725..496N,2013arXiv1307.2638N,2018Natur.562..545C,2018PhRvL.121b1303V,2019PhRvD.100j3523M,2019PhRvL.122f1105F,2019ApJ...876L...7S}. We discuss prospects for measurements of $H_0$ using gravitational waves in more detail in Section \ref{subsec:H0}.

\begin{figure*}
\centering
\includegraphics{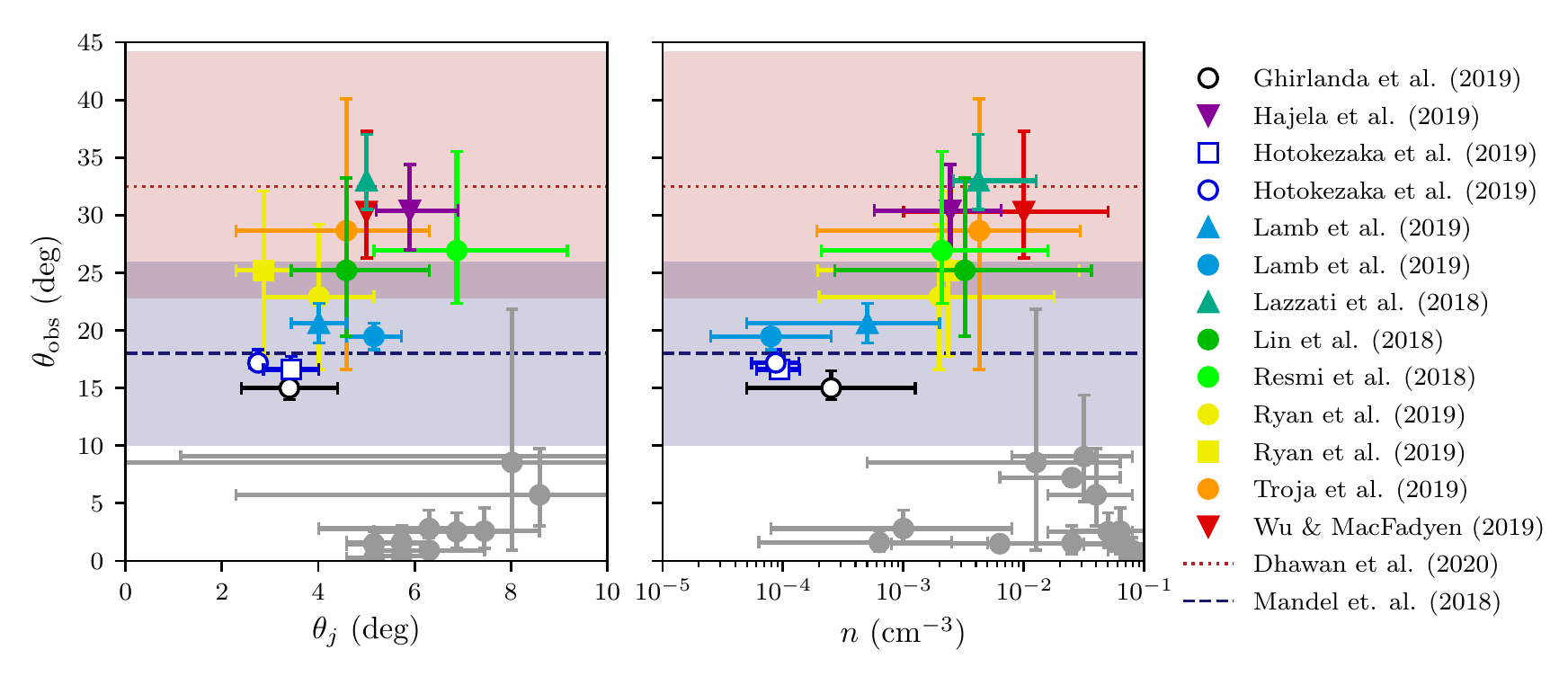}
\caption{Estimates of the observing angle ($\theta_{\rm obs}$) of GW170817 as a function of jet opening angle ($\theta_j$) and circum-merger density ($n$) for a range of Gaussian jet (circles), power-law jet (squares), boosted fireball (down triangles) and structured jet (up triangles) afterglow models. Open markers denote measurements that incorporate measurements of centroid motion, closed markers are standard lightcurve modelling, and horizontal lines correspond to estimates of $\theta_{\rm obs}$ ($1\sigma$ uncertainty shaded) that are independent of the non-thermal afterglow. Grey markers represent parameter estimates for previously observed short gamma ray bursts \citep{2019ApJ...880L..23W}.}
\label{fig:lit_comparison}
\end{figure*}

The broadband radio lightcurve of GW170817 was sampled at a cadence that was sufficient to constrain the spectral and temporal behaviour of the source. Depending on the neutron star merger detection rates in future observing runs with the Laser Interferometer Gravitational-Wave Observatory (LIGO) and Virgo detectors \citep{2018LRR....21....3A}, it may not be possible to perform radio monitoring at a similar high cadence for future events, which may result in an under\changes{sampled (and possibly unconstraining)} lightcurve. Additionally, some events may not be localised until hundreds of days post-merger if detected via radio emission alone \citep{2019PASA...36...19D}, in which case the early-time behaviour of the source will be unknown.

Even for events with a well-sampled lightcurve the information obtained about the properties of the jet and the surrounding environment is somewhat degenerate \changes{\citep{2018MNRAS.478..407N}}, \changes{although it may be possible to infer the qualitative merger geometry \citep[e.g.][]{2018MNRAS.478L..18T,2018MNRAS.481.2581L}. However, techniques like VLBI and polarisation measurements will be important in understanding and tightly constraining merger energetics and outflow geometry quantitatively}. The angular size, and therefore physical size, of the source may also be measured through observations of interstellar scintillation \citep{Goodman1997} which has previously been used  to constrain the size of gamma-ray burst (GRB) outflows \citep[e.g.][]{1997Natur.389..261F,2008ApJ...683..924C}. Understanding the size of afterglows will be an important factor in understanding their physical behaviour \citep[e.g.][]{2018MNRAS.481.2711G,2018PhRvL.120x1103L,2018MNRAS.478..407N}, and may place constraints on merger inclination in scenarios where VLBI cannot.

In this paper we discuss the detectability of expansion and motion of outflow from neutron star mergers through VLBI and observations of interstellar scintillation \changes{and implications for understanding the neutron star merger population and constraining the Hubble Constant.}

\section{The geometry of radio afterglows}
\label{sec:geometry_models}
The observed size and motion of the  afterglow of a relativistic blast wave have been studied in the literature (e.g. \citealt{Sari1998,Granot1999,Gill2018,2018Natur.561..355M,Xie2018,2019NatAs...3..940H}). For instance, the observed  size of a relativistic blast wave seen by an on-axis observer is $\sim \alpha \Gamma c T$, where $\Gamma$ is the Lorentz factor of the blast wave, $T$ is the observer time, and $\alpha$ is a numerical factor determined by the outflow's dynamics. For an on-axis blast wave   decelerating in the ISM the size is analytically given by \citet{Sari1998};
\begin{equation}
\theta_{\rm S} \approx 20\,{\rm \mu as}\,\left(\frac{n}{1\,{\rm cm}^{-3}}\right)^{-1/8}
\left(\frac{E_{\rm iso}}{10^{52}\,{\rm erg}}\right)^{1/8}
\left(\frac{T}{1\,{\rm day}}\right)^{5/8} \left(\frac{d}{100\,{\rm Mpc}}\right)^{-1}, 
\end{equation}
where $n$ is the ISM density, $E_{\rm iso}$ is the isotropic-equivalent kinetic energy,
and $d$ is the distance to the source. This estimate is valid for $\Gamma\gtrsim 1/\theta_j,\,1/\theta_{v}$, where $\theta_j$ is the jet half-opening angle and  $\theta_v$ is the viewing angle from the jet axis.

Unlike gamma-ray bursts which are preferentially seen on-axis because of Doppler beaming, afterglows of gravitational-wave mergers are most likely to be seen from a direction far from the axis of a collimated jet. In fact, \changes{by fitting both spatial information from VLBI observations and the afterglow lightcurve,} \citet{2018Natur.561..355M} find that the jet in GW170817 was observed with a viewing angle of $\theta_{v}\approx 15\degree$--$25\degree$ and estimate the jet half-opening angle as $\theta_{j}\lesssim 5\degree$. \changes{These estimates are comparable to values inferred by independent measurements using a variety of models of the non-thermal afterglow (see Table \ref{tab:lit_comparison}), although we note that these fits generally infer somewhat larger viewing angles. Additionally, \citet{2020ApJ...888...67D} find $\theta_v = 32.5_{-9.7}^{+11.7}\degree$ using thermal afterglow modelling, while \citet{2018ApJ...853L..12M} estimates $\theta_v = 18\pm 8\degree$ independent of the merger afterglow. Figure \ref{fig:lit_comparison} shows the parameter space occupied by these estimates in comparison to previously observed short GRBs.}

\changes{When observed from these angles} the flux center of the radio afterglow exhibits motion perpendicular to the line of sight and the apparent velocity may be faster than the speed of light. The afterglow light curve arising from a decelerating jet typically peaks when $\theta_v\approx 1/\Gamma$. The apparent velocity at the peak is estimated as
\begin{equation}
\beta_{\rm app,max} \approx \Gamma \beta
\end{equation}
corresponding to
\begin{equation}
\approx 1.7{\rm cot \theta_v}\,{\rm \mu as/day}\,\left(\frac{d}{100\,{\rm Mpc}}\right)^{-1},
\end{equation}
where we assume that the  opening angle of the emitting region is much less than the viewing angle.

The observed size and motion may be different from the above estimates when an outflow is seen from off-axis and the outflow has some structure , i.e., structured jets. \changes{A structured jet arises from the interaction of the jet with the merger ejetcta and the central engine activity \citep{2018MNRAS.479..588G,2018ApJ...863...58X,2018PhRvL.120x1103L}. Structured jets are often modeled by using a simple analytic function, e.g., a power law or Gaussian function. 
In this work, motivated by the  afterglow observations of GW170817, we calculate the synchrotron radio flux, centroid motion, and source size assuming  a structured jet model described by a power-law function for the angular distribution of kinetic energy}
\begin{equation}
E(\theta) = \frac{E_{\rm iso}}{1+(\theta/\theta_{j,c})^{3.5}},
\end{equation}
where $\theta$ is the polar angle from the jet axis, $\theta_{j,c}$ and $E_{\rm iso}$ are the half opening angle  and isotropic-equivalent energy of the  core of a jet.
The initial Lorentz factor of a jet is also assumed to have a power-law distribution
\begin{equation}
\Gamma(\theta) = 1+\frac{\Gamma_{c}}{1+(\theta/\theta_{j,c})^5},
\end{equation}
where $\Gamma_c$ is the Lorentz factor of the jet's core. With these distributions, we solve the radial expansion of a jet in a uniform ISM density. 
In the following, we use $\theta_{j,c}=0.05\,{\rm rad}$, $E_{\rm iso}=10^{52}\,{\rm erg}$, and $\Gamma_c=600$, for which  the afterglow light curve and superluminal motion are consistent with the observed data of GW170817 \citep{2019NatAs...3..940H}.
With these jet dynamics, we calculate the radio flux arising from the shock produced by a jet expanding in the ISM with the standard synchrotron afterglow model \citep{1998ApJ...497L..17S}. \changes{We note that our results depend only weakly on the choice of the analytic function of the structure as long as the properties of the jet core is fixed.}
We set the fraction of the shock energy that goes into the electrons and the magnetic field to be $\epsilon_e=0.1$ and $\epsilon_B=0.01$ respectively, and set the electron energy distribution to a power law with index $p=2.16$, comparable to GW170817 \citep{2018ApJ...868L..11M,2019ApJ...886L..17H}. 

We assume that radio observations occur at frequencies above the characteristic frequency of the slowest electrons, and below the synchrotron cooling break, such that the radio spectrum is reasonably modeled with a single (negatively sloped) power-law component. We set the spectral index to $\alpha=-0.585$ as observed in GW170817 \citep{2018ApJ...863L..18A,2018ApJ...868L..11M,2019MNRAS.489.1919T}.

In our afterglow models we use a 2D Cartesian coordinate system with origin at the position of the merger and define the $y$ coordinate to be along the merger axis. To determine the size and position of the afterglow emission region we define the flux weighted quantity
\begin{equation}
\langle Q \rangle = \frac{\int Q\nu dS}{\int \nu dS}
\end{equation}
where $S$ is the flux at the position and $\nu$ is the frequency. The centroid of the emission region is then given by $\langle x \rangle$, and the size of the emission region in the $x$ and $y$ directions is $\sqrt{\langle x \rangle^2 - \langle x^2 \rangle}$ and $\sqrt{\langle y^2 \rangle}$ respectively.

\changes{To ensure that this work can be generalised to all future events, we also provide scaling relations for all relevant quantities in terms of the energetics and microphysics parameters. The flux density scales according to
\begin{equation}
    S \propto E_{\rm iso}(\epsilon_{B}n)^{(p+1)/4}\epsilon_{e}^{p-1},
\end{equation}
while the size and centroid of the emission and the time since merger region scale as $(E_{\rm iso}/n)^{1/3}$. We consider values of $E_{\rm iso}=4\times 10^{49},\, 1.8\times 10^{51},\, 4.5\times 10^{52}$\,erg, corresponding to the range and median isotropic equivalent gamma ray energy of short GRBs \citep{2015ApJ...815..102F}. We consider $\epsilon_{e}=0.01,\,0.3,\, 0.5$ and $\epsilon_{B}=10^{-4},\, 2\times 10^{-3},\, 2\times 10^{-2}$ based on the estimates and 1-$\sigma$ uncertainties of these parameters for GW170817 \citep{2019ApJ...886L..17H}.
}

\begin{figure}
\includegraphics[width=\linewidth]{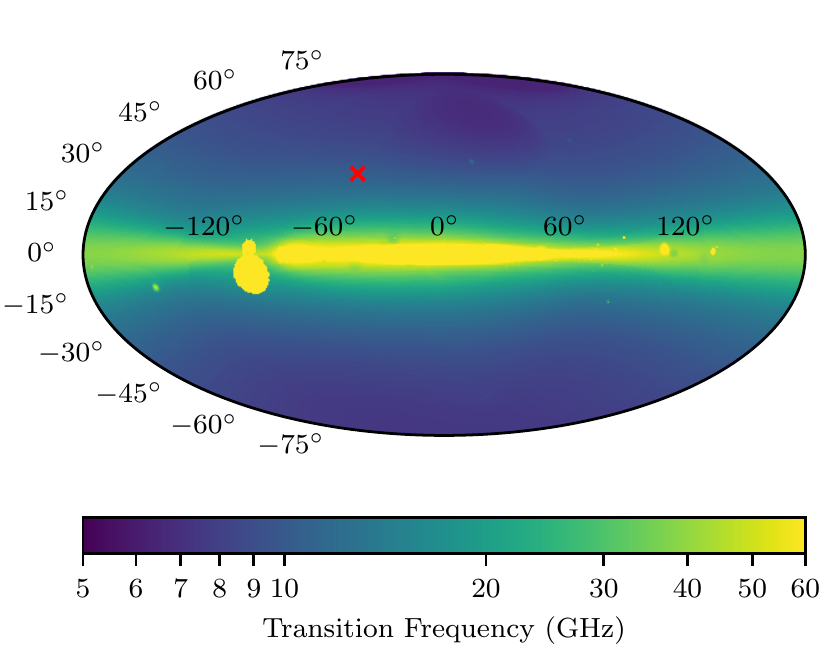}
\caption{Transition frequency as a function of Galactic coordinates for extragalactic sources. The red cross denotes the location of GW170817.}
\label{fig:transition_frequency_allsky}
\end{figure}

\section{Observations of interstellar scintillation}
Interstellar scintillation is the observed extrinsic variability of radio sources induced by inhomogeneities in the electron density along their line of sight \citep{1995ApJ...443..209A}. The induced variability of a radio source can be quantified with the scattering strength, given by
\begin{equation}
    \xi=2.6\times 10^{3} {\rm SM}^{0.6}D^{0.5}\nu^{-1.7}
\end{equation}
where ${\rm SM}$ is the scattering measure (which describes the cumulative contribution of the inhomogeneities along the line of sight), $D$ is the distance to the equivalent phase screen in kpc, and $\nu$ is the observing frequency in GHz \citep{1992RSPTA.341..151N,1993ApJ...411..674T,1998MNRAS.294..307W}.

\citet{1998MNRAS.294..307W,2001MNRAS.321..176W} describes scaling relations for various scintillation parameters at an observing frequency, $\nu$, as a function of the transition frequency, $\nu_0$, the frequency at which the scattering strength is unity, and the size of the Fresnel zone at the transition frequency given by $\theta_{\rm F0}=8/\sqrt{D\nu_0}$ and the observing frequency. The scattering strength itself depends on the distribution of electrons along the line of sight, which we assume is dominated by electrons within the Milky Way for simplicity. The variability timescale is dependent on the transverse speed of the phase screen relative to the observer and source across the line of sight, and the spatial distribution of electrons along the line of sight. We adopt the same approximation as \citet{1998MNRAS.294..307W} and assume that the variability timescale at the transition frequency is 2\,hours in all directions based on a phase screen transverse velocity of $50\,$kms$^{-1}$ \citep{1995A&A...293..479R}.

Scintillation occurs in two main regimes; weak scattering ($\nu>\nu_0$, where the phase changes in the observed radio signal introduced by the interstellar medium are negligible) and strong scattering ($\nu<\nu_0$, where the phase changes are the dominant cause of the observed variability). Additionally, there are two forms of strong scattering; refractive scintillation which is characterised by broadband variability on timescales of days, and diffractive scintillation which is characterised by narrow-band variability on much shorter timescales. Scintillation is typically observable in sources that are smaller than the size of the relevant scattering disk ($\theta_{\rm F}$ for weak scattering, and some multiple of $\theta_{\rm F}$ for strong scattering), although \citet{1992RSPTA.341..151N} provides scaling relations for sources that are larger than the scattering disk.

We use NE2001 \citep{2002astro.ph..7156C,2003astro.ph..1598C}, a model for the Galactic electron distribution based on independent measurements of the dispersion measure (DM) and distance of 269 pulsars, to retrieve values of $\nu_0$ and $\theta_{\rm F0}$ along various lines of sight. NE2001 has been widely used for scattering calculations in the past, but other electron density models including YMW16 \citep{2017ApJ...835...29Y}, which uses similar techniques to NE2001, and RISS19 \citep{2019arXiv190708395H}, which uses H$_\alpha$ measurements to trace free electrons, may also be suitable and will produce qualitatively similar results. \changes{We note that both NE2001 and YMW16 only consider the contribution of Galactic electrons, and therefore an extragalactic measure like RISS19 should provide more accurate estimates. However, RISS19 is less tested against observations than other models and therefore we choose to use NE2001 for easy comparison to the literature}

\changes{The analysis in the following sections focuses on the synchrotron emission produced by the forward shock of the merger outflow, however at very early times ($t\lesssim 1\,$day), when the outflow is extremely compact, the radio emission may be dominated by the reverse shock. See \citet{2019MNRAS.489.1820L} for a discussion of the effects of scintillation on emission from the reverse shock.}

\subsection{Prospects for detection of scintillation}
We perform a qualitative analysis of the detectability of all forms of scintillation at a range of frequencies. We divide the radio spectrum into four parts: low frequencies ($\sim$300\,MHz), mid frequencies (0.8--2\,GHz), high frequencies (2--10\,GHz) and millimetre wavelengths (>20\,GHz) based on the capabilities of existing radio facilities. In general, scintillation is easier to detect in sources that have small angular sizes, i.e. more distant, off-axis, events. However, in the case of relativistic outflows both of these properties correspond to lower observed flux densities, and therefore make the emission and any variability more difficult to detect. Events occuring in dense environments are more luminous and remain compact for longer.

\begin{figure}
\includegraphics{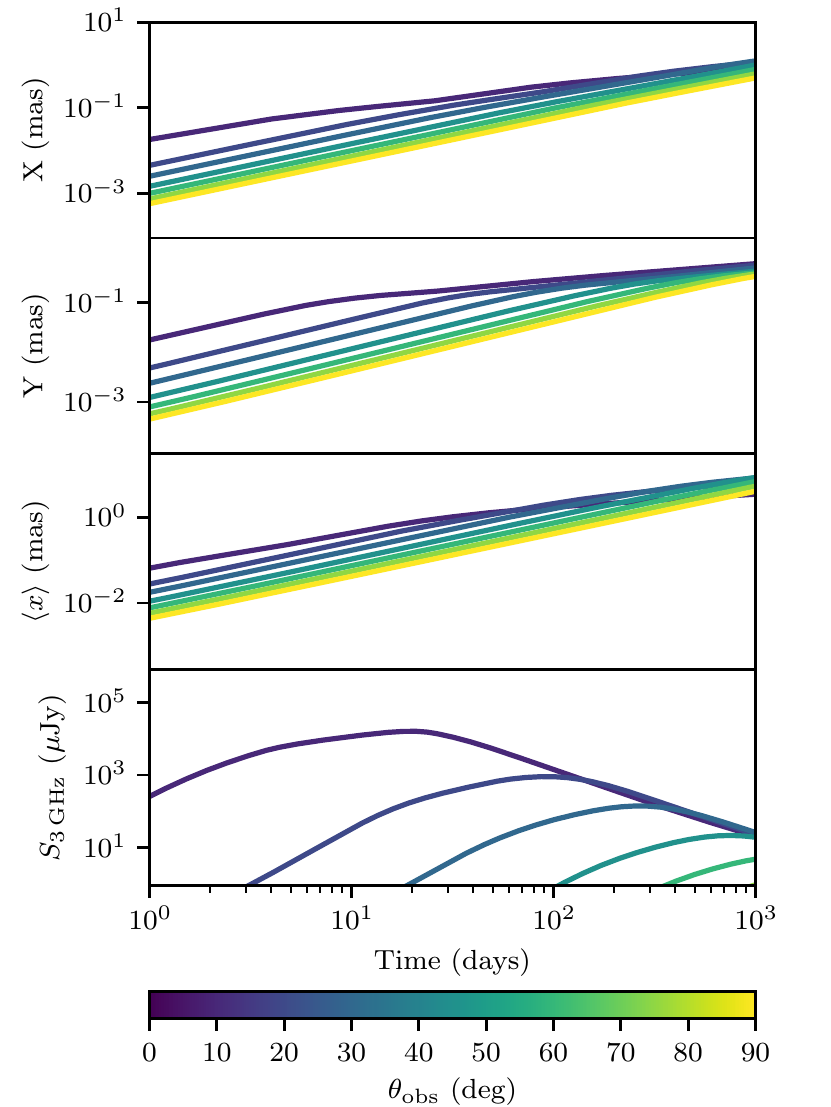}
\caption{Angular size (top 2 rows), centroid offset (row 3) and 3\,GHz flux density (bottom) for a neutron star merger \changes{jet model} at a distance of 40\,Mpc with a circum-merger density of $n=10^{-3}\,$cm\changes{$^{-3}$} for a range of inclination angles.}
\label{fig:motion+flux}
\end{figure}

\subsubsection{Weak scattering}
At observing frequencies $\nu > \nu_0$ (or equivalently $\xi < 1$) we will observe weak scattering where the dominant cause of phase changes is path-length variations.

The size of the scattering disk (which determines whether an object can be treated as a point source) is simply the Fresnel zone, $\theta_{\rm F}$, which at the transition frequency is $<5\,\mu$as for all positions on the sky ($\theta_{\rm F0}<0.5\,\mu$as near the Galactic plane), and scales as $\nu^{-1/2}$. This is typically smaller than the angular size of the source, $\theta_{\rm s}$, by the time it becomes luminous enough to be detectable (e.g. see Figure \ref{fig:motion+flux}). The modulation index of weak scattering is strongly dependent on observing frequency, and is given by
\begin{equation}
    m_{\rm W} = \left(\frac{\nu_0}{\nu}\right)^{17/12}\min\left[1,\frac{\theta_{\rm F}}{\theta_{\rm s}}\right]^{7/6}
\end{equation}
Hence, as we move above the transition frequency and further into the weak scattering regime both the modulation index and size of the Fresnel zone decrease.

Additionally, early-time observations at frequencies above the transition frequency are less feasible compared to those at lower frequencies due to the negative spectral index of the afterglow. Combining these three factors we find that detecting variability due to weak scattering from radio afterglows will be quite difficult with current and planned radio telescopes and do not consider it any further.

\begin{table}
	\centering
	\caption{Specifications of radio facilities we consider in the detection of refractive and diffractive scintillation. $\nu_{\rm obs}$ is the center observing frequency, BW is bandwidth, $t_{\rm obs}$ is the proposed observing time and $\sigma_{\rm min}$ is the corresponding image sensitivity. \changes{For existing facilities we use noise estimates based on the achieved sensitivity in observations of GW170817 \citep[e.g.][]{2018ApJ...868L..11M}, taking into account increased noise do to host galaxy emission, while we use thermal noise estimates for the SKA \citep{2019arXiv191212699B}.}}
	\label{tab:facility_specs}
	\begin{tabular}{lcccc}
    	\hline\hline
    	Telescope & $\nu_{\rm obs}$ & BW & $t_{\rm obs}$ & $\sigma_{\rm min}$\\
    	~ & (GHz) & (GHz) & (hours) & ($\mu$Jy)\\
    	\hline
    	VLA-S & 3.0 & 1.5 & 3 & 2.4\\
    	VLA-C & 6.0 & 4.0 & 3 & 1.5\\
    	ATCA-CX & 7.2 & 4.0 & 12 & 8\\
    	\hline
    	SKA-1 Mid & 1.4 & 0.77 & 3 & 0.45\\
    \hline
    \end{tabular}
\end{table}

\subsubsection{Diffractive Interstellar Scintillation (DISS)}
\label{subsec:diss_description}
For observing frequencies $\nu < \nu_0$ we are in the strong scattering regime, where two forms of scintillation are present. Diffractive scintillation is fast (minutes--hours), narrow-band variability typically observed in compact Galactic objects like pulsars \citep[e.g.][]{1977ARA&A..15..479R,1986ApJ...311..183C,1992Natur.360..137P,1997NewA....2..449G,2016MNRAS.461..908B}. We must consider the detectability of such variations on the small frequency- and time-scales involved, which require consideration of the telescope sensitivities on those scales. For optimal detectability we require our time resolution to be less than the variability timescale and the channel width to to be comparable to the frequency scale, which  is given by
\begin{equation}
\Delta\nu = \nu \left(\frac{\nu}{\nu_0}\right)^{17/5}.
\label{eq:diffractive_freq_scale}
\end{equation}
The size of the scattering disk is given by
\begin{equation}
\theta_{d} = \theta_{F0} \left(\frac{\nu}{\nu_0} \right)^{6/5},
\label{eq:scattering_disk_diffractive}
\end{equation}

Low frequencies ($\leq 300$\,MHz) are in the strong scattering regime for the entirety of the sky. However, the diffractive scintillation bandwidth for compact extragalactic sources at low frequency is typically $\sim 3\,$kHz, smaller than the continuum channel width of current generation telescopes at this frequency. 

The scintillation timescale of diffractive scintillation is given by
\begin{equation}
t_d = 2\,{\rm hour}\left(\frac{\nu}{\nu_0} \right)^{6/5} \max\left[1,\frac{\theta_{\rm S}}{\theta_{d}} \right],
\label{eq:timescale_diffractive}
\end{equation}
and in this situation is typically 1 minute. Therefore detecting DISS requires sub-minute observations split into single channels. The estimated sensitivity for this type of observation with current generation low-frequency telescopes is on the order of hundreds of mJy. In comparison, the flux density of GW170817 would have been $<1\,$mJy at its peak.

The size of the scattering disk at low frequencies is significantly smaller than the Fresnel zone, meaning that any variability will have a low amplitude since the modulation index of diffractive scintillation is given by
\begin{equation}
m_d = \min\left[1, \frac{\theta_d}{\theta_{\rm S}}\right].
\label{eq:mod_index_diffractive}
\end{equation}
Therefore, with the exception of bright events, DISS from gravitational wave afterglows will not be detectable at low frequencies with current radio facilities.

Frequencies around 1\,GHz accessible with telescopes such as the Australian Square Kilometre Array Pathfinder \citep[ASKAP;][]{2008ExA....22..151J}, the Karl G. Jansky Very Large Array (VLA) and MeerKAT\footnote{\url{http://www.ska.ac.za/gallery/meerkat/}} are in the strong scattering regime for the entirety of the sky. The typical diffractive scintillation timescale at these frequencies is a few minutes. \citet{2019MNRAS.488..559Z} demonstrate the capability of ASKAP to detect short-timescale variability with high spectral resolution by performing dynamic spectroscopy of UV Ceti (a bright, well-known flare star), achieving a sensitivity equivalent to $\sim 12$\,mJy in a 10 second integration is per 1\,MHz channel. However, current telescopes do not have the instantaneous sensitivity required to detect scintillation from sources at distances comparable to the LIGO horizon, which will likely peak below 1\,mJy in this frequency regime \citep{2019PASA...36...19D}.

Additionally, the characteristic frequency scale of diffractive scintillation at these frequencies is $< 1$\,MHz across the sky. Current generation gigahertz-frequency telescopes have channel widths of 1\,MHz and therefore do not have sufficient spectral resolution to detect extragalactic diffractive scintillation.

Frequencies up to 10\,GHz are in the strong scattering regime away from the Galactic poles ($|b|\lesssim 40\degree$). At 3\,GHz typical values for the scintillation characteristic bandwidth and timescale are tens of MHz and tens of minutes respectively. The sensitivity of a 5 minute observation with the VLA at 3\,GHz using 5\,MHz of bandwidth is $300\,\mu$Jy. At frequencies closer to 10\,GHz the characteristic bandwidth and timescale of diffractive scintillation both increase to values of $\sim 1\,$GHz and $\sim 1$ hour. Both of these represent reasonable prospects of detecting scintillation, and we perform a more quantitative analysis in Section \ref{subsec:detect_diffractive}.

Millimetre wavelengths are only in the strong scattering regime close to the Galactic plane, with characteristic bandwidths of a few GHz and timescales comparable to those at lower frequencies. \changes{We expect that the spectral index of the radio afterglow will be negative \citep{1998ApJ...497L..17S,2014ARA&A..52...43B}, although we note that this may not be true in all cases. Therefore in general,} sources will be more difficult to detect at these frequencies compared to observations at lower frequencies. However, emission from mergers occuring in environments that are more dense than the typical short GRB circum-burst density \changes{\citep[$n\sim 10^{-2}$\,cm$^{-3}$;][]{2015ApJ...815..102F}}, may be detectable with relatively short integrations. Therefore we do not perform any further analysis, but do not rule out the possibility of detecting DISS with millimetre observations.

\subsubsection{Refractive Interstellar Scintillation (RISS)}
Refractive scintillation manifests as slower broadband changes and is observed in pulsars as well as compact extragalactic sources like quasars. For refractive interstellar scintillation the size of the scattering disk, $\theta_r$, is given by
\begin{equation}
\theta_{r} = \theta_{F0} \left(\frac{\nu_0}{\nu} \right)^{11/5},
\label{eq:scattering_disk_refractive}
\end{equation}
the modulation index, $m_r$, is given by
\begin{equation}
m_r = \left(\frac{\nu}{\nu_0} \right)^{17/30} \min\left[1,\left(\frac{\theta_{r}}{\theta_{\rm S}} \right)^{7/6} \right],
\label{eq:mod_index_refractive}
\end{equation}
and the variability timescale, $t_r$, is given by
\begin{equation}
t_r = 2\,{\rm hour}\left(\frac{\nu_0}{\nu} \right)^{11/5} \max\left[1,\frac{\theta_{\rm S}}{\theta_{r}} \right].
\label{eq:timescale_refractive}
\end{equation}

The modulation index for RISS is typically <10\% for a compact source at low frequencies, and even lower for sources larger than the size of the scattering disk. We therefore do not expect to detect any form of interstellar scintillation from GW radio afterglows with current low frequency telescopes.

Away from the Galactic plane the modulation index for refractive scintillation from compact sources ranges from 0.3--0.45 for mid-frequencies, with a typical timescale of a few days. The size of the scattering disk is $\theta_r\approx 0.5$\,mas for $\theta_{F0}\sim3\,$mas and $\nu_0\sim10\,$GHz (typical values away from the Galactic plane), meaning that the radio afterglow behaves as a compact source until late times. At high frequencies the modulation index for compact sources is larger, but the size of the scattering disk is smaller and therefore the point-source approximation does not apply for as long. We investigate prospects for detecting RISS at mid and high frequencies in Section \ref{subsec:detect_refractive}

At mm-wavelengths, the timescale of refractive scintillation is comparable to $t_d$. For similar reasons as in Section \ref{subsec:diss_description} we do not perform further analysis, but do not rule out the possibility of detecting RISS with mm observations for events near the Galactic plane.

\begin{figure*}
\centering
\includegraphics{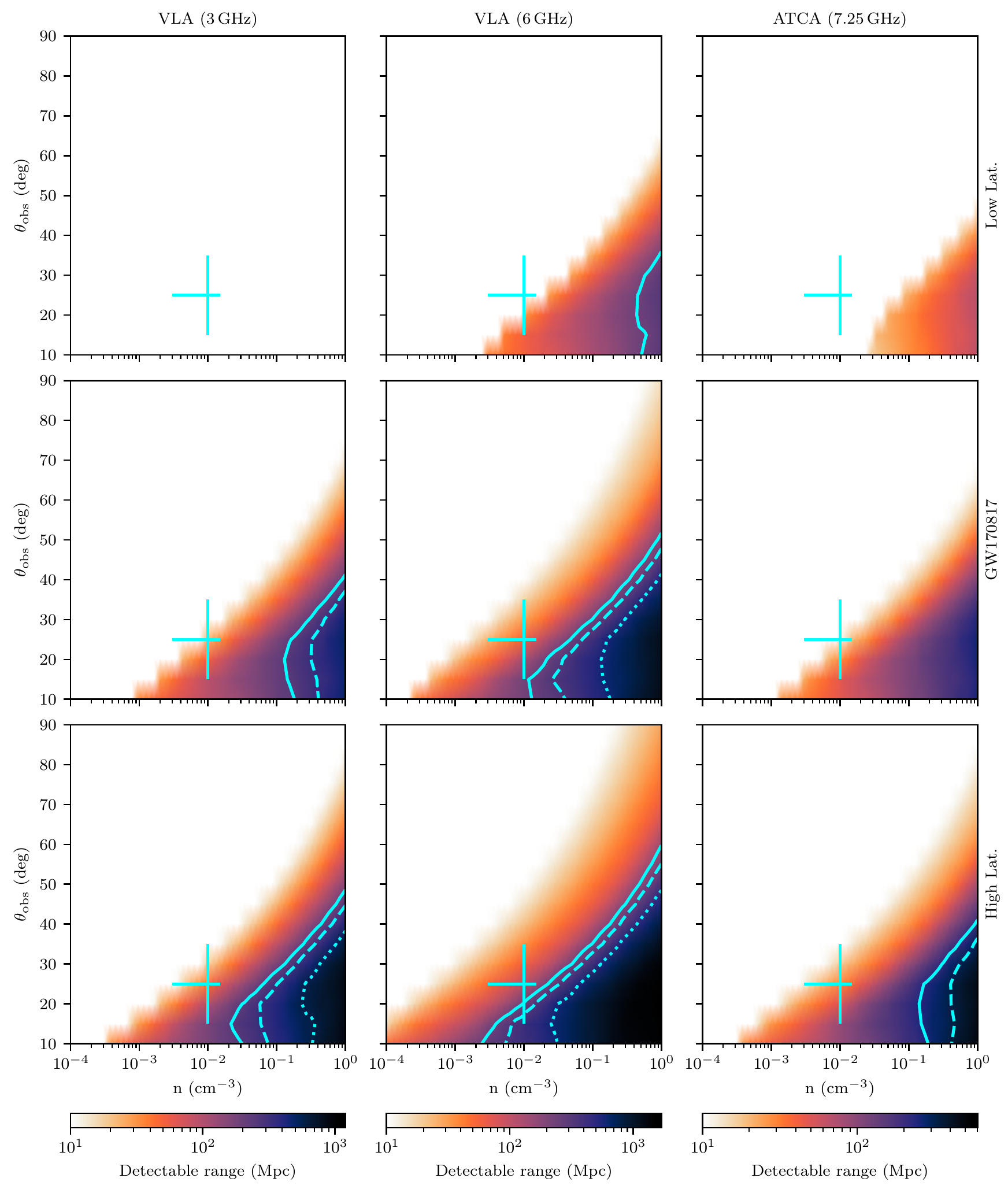}
\vspace{-20pt}
\caption{Maximum distance at which diffractive scintillation is detectable for a range of electron density parameters. Top: typical scintillation parameters at low Galactic latitudes ($10\degree<|b|<20\degree$), $\nu_0 = 18.1$\,GHz and $\theta_{F0} = 1.5\,\mu$as. Middle: $\nu_0 = 10.3$\,GHz and $\theta_{F0} = 2.9\,\mu$as corresponding to the line of sight to GW170817. Bottom: typical scintillation parameters at high Galactic latitudes ($60\degree<|b|<70\degree$), $\nu_0 = 7.82$\,GHz and $\theta_{F0} = 3.9\,\mu$as. This is shown for observations with the VLA at 3\,GHz (left, not detectable for low $|b|$), 6\,GHz (middle) and observations with the ATCA at 7.25\,GHZ (right). The jagged edge is an artefact of using simulating models with steps in inclination angle of $5\,deg$. Contours corresponding to the inclination angle dependent LIGO horizon for O3 (135\,Mpc, solid), design specifications (190\,Mpc, dashed), \changes{and A+ (330\,Mpc, dotted) are shown in blue. The blue cross corresponds to the typical short GRB circum-merger density and estimates for the inclination angle of GW170817.}}
\label{fig:diffscintillation_distance}
\end{figure*}

\subsection{A Generalised Metric for Detecting Scintillation}
Our ability to detect variability is strongly dependent on the exact observing strategy, most notably the total number of observations, observation sensitivity and observing cadence. We therefore define a generalised detectability metric that allows us to place broad estimates on the range at which scintillation may be detected for a combination of merger parameters. We emphasise that this metric should not be used in preparing follow-up observations of individual events and instead more detailed, event-specific calculations should be performed. 

For scintillation to be detectable we require $m > m_{\rm detect}$ and $S > S_{\rm detect}$ where $m$ is the modulation index of the source due to scintillation, $S$ is the flux density, while $m_{\rm detect}$ and $S_{\rm detect}$ are the minimum detectable values for each of those quantities respectively. We define both detectability thresholds in terms of the image RMS, $\sigma$; $S_{\rm detect} = 5\sigma$ and $m_{\rm detect} = 5\sigma / S$. For some events scintillation may be detected on timescales of days--weeks in the form of inter-observation variability, but other events may exhibit intra-observation variability. As such we define

\begin{equation}
    \sigma =
    \begin{cases}
    \sigma_{\rm min} & \text{for }  t_{\rm obs} \geq t_s\\
    \sigma_{\rm min}\sqrt{t_{\rm obs}/t_s} & \text{for }  t_{\rm obs} < t_s
    \end{cases},
    \label{eq:image_rms}
\end{equation}

where $\sigma_{\rm min}$ is the minimum reasonable image RMS achievable with a telescope, requiring an observation time of $t_{\rm obs}$ and $t_s$ is the timescale of the scintillation in question.

In the case of diffractive scintillation we also have constraints based on the scintillation bandwidth as well as the telescope channel width, frequency and bandwidth. For scintillation to be detectable we require 10 samples across the scintillation bandwidth and define the effective bandwidth as $\Delta\nu_{\rm eff}=\Delta\nu/10$. We also require that the channel width (typically 1\,MHz) for the telescope is less than $\Delta\nu_{\rm eff}$. We then correct the image RMS defined in \eqref{eq:image_rms} for the fractional bandwidth, scaling it as $\Delta\nu_{\rm eff}^{-1/2}$.

Finally, we require that the above conditions are satisfied for at least thirty days (since observers have minimal a priori knowledge of when scintillation will be detectable) and ten scintillation timescales (to allow for a sufficient number of observations to characterise the variability as being produced by scintillation).

Variability caused by scintillation is more easily detectable for sources with small angular sizes and large flux densities. More distant events have smaller angular sizes (scaling as $D^{-1}$) and lower flux densities (scaling as $D^{-2}$) and therefore scintillation is detectable for a range of distances, and not simply out to a horizon distance. However, the minimum of that range is typically $<10$\,Mpc so for brevity we simply quote the maximum detectable distance as our detectability metric due to the expected low rate of events occuring at such small distances \citep{2019PhRvX...9c1040A}.

\subsection{Scintillation Detectability}
\subsubsection{Detectability of Diffractive Scintillation}
\label{subsec:detect_diffractive}
Here we consider follow-up with 3 telescope configurations:
\begin{enumerate}
    \item VLA follow-up in S band (3\,GHz) with $t_{\rm obs} = 3$\,hours and $\sigma_{\rm min} = 2.4$\,$\mu$Jy;
    \item VLA follow-up in C band (6 GHz) with $t_{\rm obs} = 3$\,hours, $\sigma_{\rm min} = 1.5$\,$\mu$Jy;
    \item ATCA follow-up in the CX band (7.25\,GHz), with $t_{\rm obs} = 12$\,hours and $\sigma_{\rm min} = 8$\,$\mu$Jy
\end{enumerate}

The stated sensitivity reflects typical values achieved during the follow-up of GW170817 \citep{2018ApJ...868L..11M}, which was limited by radio emission from the host galaxy. It may be possible to achieve better sensitivity for events with a less luminous host galaxy.

Figure \ref{fig:diffscintillation_distance} shows the maximum detectable distance at which scintillation is detectable with each of these telescope configurations as a function of circum-merger density and observing angle for a range of scintillation parameters corresponding to a range of Galactic latitudes. As expected, on-axis events in denser environments are detectable to a larger distance as they have higher flux densities but also remain compact for longer. Almost half of the parameter space (low density, off-axis events) is inaccessible with current radio facilities. We find that away from the Galactic plane ($|b| > 30\degree$) the dependence of scintillation detectability on Galactic latitude is minimal, but not negligible.

We now compare the maximum detectable distance to the LIGO detector horizons which are averaged across the sky and inclination angles. The best detector horizon achieved to-date in O3 is 135\,Mpc, while the expected range for O3 and design sensitivity is 150 and 190\,Mpc respectively \citep{2018LRR....21....3A}. \changes{Additionally, the planned A+ upgrade that will be online for the O5 run scheduled in 2025 will increase the detector horizon to 330\,Mpc \citep{2018LRR....21....3A}.}

To find the dependence of the horizon on inclination angle we average equation (3.31) from \citet{1993PhRvD..47.2198F} across the antenna patterns. We find
\begin{equation}
    \mathcal{R}(\theta_{\rm obs}) \approx 0.658\overline{\mathcal{R}}\sqrt{1+6\cos^2\theta_{\rm obs}+\cos^4\theta_{\rm obs}}
    \label{eq:gw_inc_angle}
\end{equation}
where $\mathcal{R}(\theta_{\rm obs})$ is the inclination angle dependent range and $\overline{\mathcal{R}}$ is the quoted LIGO horizon.

The detectable range of diffractive scintillation with the VLA extends beyond the LIGO horizon for dense, on-axis events away from the Galactic plane, while the range of the ATCA is typically 3--5 times lower. Events occurring in similar environments to GW170817 \changes{($n\sim 10^{-3}$\,cm$^{-3}$; see Figure \ref{fig:lit_comparison})} will not exhibit diffractive scintillation detectable with any current radio telescopes. However, events with circum-merger densities comparable to typical short GRB circum-burst densities
\citep[$n\sim 10^{-2}$\,cm$^{-3}$;][]{2015ApJ...815..102F}
may exhibit scintillation detectable with the VLA C band receiver at distances of 100\,Mpc for $\theta_{\rm obs}<30\degree$.

\changes{We also run the simulation for the range of energetics and microphysics parameters stated in Section \ref{sec:geometry_models}, varying each parameter individually and keeping the remaining parameters at the fiducial value. The detectable range is most influenced by the isotropic equivalent energy, with the typical range varying by a factor of 0.002--3, although the best and worst cases are factors of 9 and 0.001. Varying $\epsilon_e$ and $\epsilon_B$ changes the detectable range by factors of 0.1-5 and 0.03--1.7 respectively. These factors are only very weakly dependent on the scintillation and telescope parameters, and any variance is negligible compared to the uncertainty in our models.}

\begin{figure*}
\centering
\includegraphics{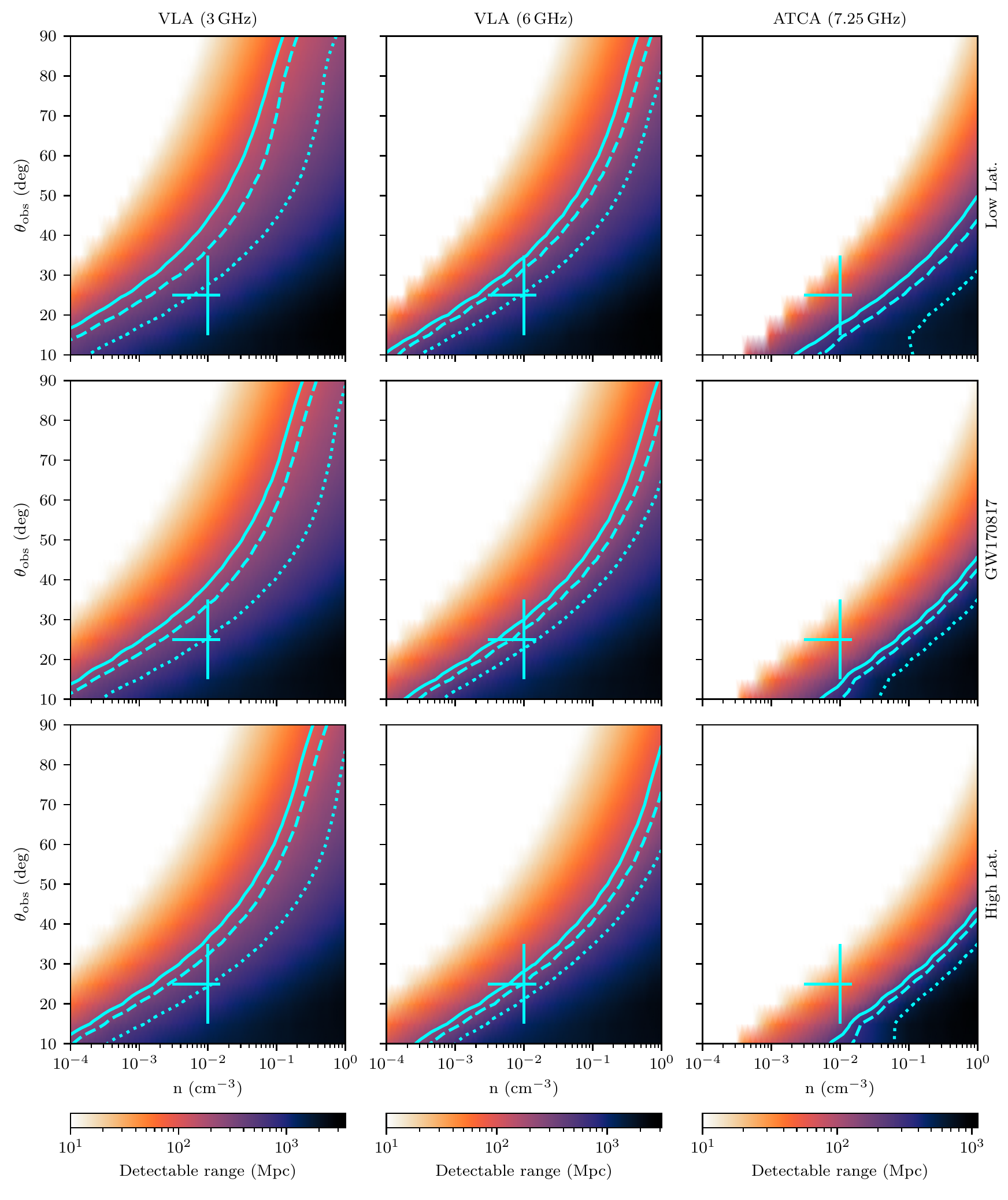}
\vspace{-20pt}
\caption{Maximum distance at which refractive scintillation is detectable for a range of electron density parameters. Top: typical scintillation parameters at low Galactic latitudes ($10\degree<|b|<20\degree$), $\nu_0 = 18.1$\,GHz and $\theta_{F0} = 1.5\,\mu$as. Middle: $\nu_0 = 10.3$\,GHz and $\theta_{F0} = 2.9\,\mu$as corresponding to the line of sight to GW170817. Bottom: typical scintillation parameters at high Galactic latitudes ($60\degree<|b|<70\degree$), $\nu_0 = 7.82$\,GHz and $\theta_{F0} = 3.9\,\mu$as. This is shown for observations with the VLA at 3\,GHz (left), 6\,GHz (middle) and observations with the ATCA at 7.25\,GHZ (right).  The jagged edge is an artefact of using simulating models with steps in inclination angle of $5\degree$. Contours corresponding to the inclination angle dependent LIGO horizon for O3 (135\,Mpc, solid), design specifications (190\,Mpc, dashed),\changes{ and A+ (330\,Mpc, dotted) are shown in blue. The blue cross corresponds to the typical short GRB circum-merger density and estimates for the inclination angle of GW170817.}}
\label{fig:scintillation_distance}
\end{figure*}

\subsubsection{Detectability of Refractive Scintillation}
\label{subsec:detect_refractive}
We consider the three observing scenarios defined above, applied to refractive scintillation. Figure \ref{fig:scintillation_distance} shows the results of this analysis. We find that the overall trend of scintillation being detectable to larger distances for events that are on-axis and occur in dense environments holds true for both forms of strong scattering. However the fraction of the $\theta_{\rm obs}$--$n$ parameter space accessible is much larger for refractive scintillation than diffractive scintillation, and the detectability range tends to be larger. We find that refractive scintillation from GW170817 may have been detectable assuming a robust, high-cadence, follow-up plan had been in place, which would have likely required knowing precise merger parameters a priori.

As in Section \ref{subsec:detect_diffractive} we run the simulation for a range of energetics and microphysics parameters. Again, the detectable range is most influenced by the isotropic equivalent energy, with the typical range varying by a factor of 0.01--2.5, although the best and worst cases are factors of 4 and 0.005. Varying $\epsilon_e$ and $\epsilon_B$ changes the detectable range by factors of 0.2--3.2 and 0.1--1.5 respectively.

\begin{figure*}
\includegraphics{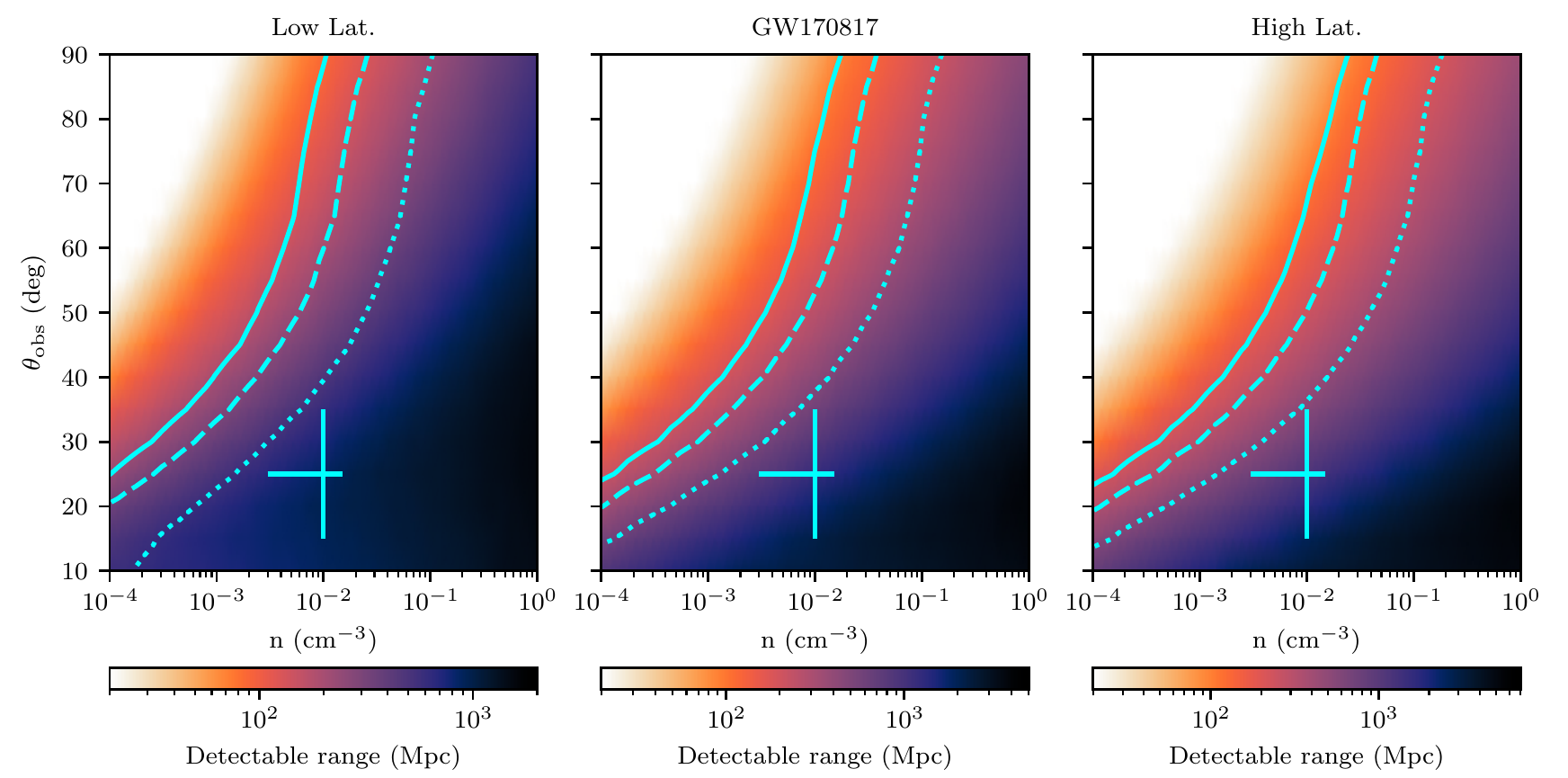}
\caption{Maximum distance at which refractive scintillation is detectable with the Square Kilometre Array for a range of electron density parameters. Left: typical scintillation parameters at low Galactic latitudes ($10\degree<|b|<20\degree$), $\nu_0 = 18.1$\,GHz and $\theta_{F0} = 1.5\,\mu$as. Middle: $\nu_0 = 10.3$\,GHz and $\theta_{F0} = 2.9\,\mu$as corresponding to the line of sight to GW170817. Right: typical scintillation parameters at high Galactic latitudes ($60\degree<|b|<70\degree$), $\nu_0 = 7.82$\,GHz and $\theta_{F0} = 3.9\,\mu$as. Contours corresponding to the inclination angle dependent LIGO horizon for O3 (135\,Mpc, solid), design specifications (190\,Mpc, dashed), \changes{and A+ (330\,Mpc, dotted) are shown in blue. The blue cross corresponds to the typical short GRB circum-merger density and estimates for the inclination angle of GW170817.}}
\label{fig:scintillation_distance_SKA}
\end{figure*}

\subsubsection{Future Prospects: Next generation radio telescopes}
\label{subsec:scint_SKA}
A significant fraction of the parameter space exhibits refractive scintillation detectable with the VLA beyond the LIGO detector horizons. However, low-density off-axis events still do not exhibit any detectable variability from either form of strong scattering. We therefore turn our focus to the Square Kilometre Array (SKA), which will have a sensitivity orders of magnitude better than existing radio telescopes. We consider observations with the SKA-1 (mid) array at $\nu=1.4\,$GHz, assuming a bandwidth of 770\,MHz and a sensitivity of 1.2\,$\mu$Jy in a 3 hour integration \citep{2019arXiv191212699B}.

Figure \ref{fig:scintillation_distance_SKA} shows that it will be possible to detect refractive scintillation from all but a small minority of low density, off-axis events detected by LIGO/Virgo with the SKA. Most events will exhibit detectable scintillation out to Gpc distances, although generally not beyond the horizon of third generation gravitational wave detectors which will come online in the 2030s-2040s \citep{2019BAAS...51g..35R} and have detection horizons of tens--hundreds of Gpc \citep{2013CQGra..30g9501S}. Like the current situation, the most dense and on-axis events will be detectable beyond the gravitational wave detector horizon.

We have also applied the criteria outlined in Section \ref{subsec:detect_diffractive} to the SKA. We find that no events will exhibit diffractive scintillation detectable with SKA continuum observations due to the scintillation bandwidth being smaller than the continuum channel width. However, we find that using spectral line observing, and assuming a channel width given by $\Delta\nu/\nu=10^{-4}$ the SKA has a detectability horizon $\sim 5$ times larger than observations in VLA C band. As well as having a larger horizon, these observations will allow tighter constraints to be placed on source sizes at early times, as the scattering disk is almost six times smaller.

The dependence on energetics and microphysics parameters for these ranges is comparable to the values in Section \ref{subsec:detect_diffractive} and \ref{subsec:detect_refractive}.

Two other major radio facilities are expected to come online on similar timescales. The 2000 antenna Deep Synoptic Array \citep[DSA;][]{2019BAAS...51g.255H} will have a 1\,hr continuum sensitivity of $1\,\mu$Jy and while it will be more suited to discovering radio emission from compact object mergers due to it's large field of view and high survey speed, it will also be capable of observing scintillation from events within $\sim 80\%$ of the estimated range of the SKA. The next-generation VLA \citep[ngVLA;][]{2019BAAS...51g..81M,2019arXiv190310589C,2019BAAS...51c.209C} will improve the sensitivity and resolution of the VLA by a factor of 10, corresponding to a detector horizon that is $\sim 3$ times larger than the existing VLA.

\begin{table}
	\centering
	\caption{Estimated parameters for 3 VLBI observing scenarios. The HSA consists of the VLBA, the phased VLA, the GBT and Arecibo (Ar). $S_{\rm noise}$ is the estimated thermal noise in the observation, $\theta_{\rm B}$ is the approximate beam size, $\Delta\alpha\cos\delta$ and $\Delta\delta$ are the systematic uncertainties in R.A. and Dec. respectively, and $\theta_{\rm sys}$ is the systematic astrometric uncertainty we use for this analysis, estimated by taking the geometric mean of $\Delta\alpha\cos\delta$ and $\Delta\delta$.}
	\label{tab:VLBI_properties}
	\begin{tabular}{lccc}
	    \hline
	    \hline
		 & LBA & HSA & HSA (no Ar)\\
		\hline
		$\nu$ (GHz) & 4.8 & 4.5 & 4.5\\
		Obs. time (h) & 12 & 2 & 8\\
		Dec ($\degree$) & -30 & 20 & 50\\
		\rule{0pt}{4ex}$S_{\rm noise}$ ($\mu$Jy) & 20 & 3.2 & 3.1\\
		$\theta_{\rm B}$ (mas) & 15 & 3 & 3\\
		\rule{0pt}{4ex}$\Delta\alpha\cos\delta$ ($\mu$as) & 80 & 60 & 80\\
		$\Delta\delta$ ($\mu$as) & 100 & 80 & 100\\
		$\theta_{\rm sys}$ ($\mu$as) & 90 & 70 & 90\\
		\hline
		\hline
	\end{tabular}
\end{table}

\section{VLBI Observations}
VLBI observations of GW170817 were important in determining the geometry of the merger, constraining both the emission model (jet-dominated) and the inclination angle ($\sim 20\degree$) via observations of a positional shift in the source centroid \citep{2018Natur.561..355M,2019Sci...363..968G}. In this section we discuss prospects for directly imaging outflow structure using high resolution VLBI imaging and expand on the work of \citet{2019A&A...631A..39D} by determining a detection metric driven by observing constraints.

The parameters of any VLBI observation are strongly dependent on the declination of the target source. Northern Hemisphere sources are accessible with the High Sensitivity Array (HSA), consisting of the Green Bank Telescope, the phased VLA, Arecibo and the Very Long Baseline Array (VLBA). The longest baseline spans 8611\,km (Mauna Kea--Saint Croix), corresponding to a best achievable angular resolution of 0.8\,mas at 9\,GHz, which is the highest frequency available to Arecibo. However, observations with Arecibo longer than 2 hours are only possible between declinations of +5 and +30\,$\degree$\footnote{\url{http://www.naic.edu/~astro/aovlbi/}}. Using the European VLBI Network (EVN) calculator\footnote{\url{http://www.evlbi.org/cgi-bin/EVNcalc}} the flux density sensitivity of the array in a 2 hour observation ranges from 2.9--3.6\,$\mu$Jy for observing bands from 1.4--9\,GHz. Observations without Arecibo allow for longer integration times, so we additionally consider an 8 hour observation, achieving a sensitivity of $3.1\,\mu$Jy.

Southern Hemisphere sources are accessible with the Long Baseline Array (LBA), consisting of the phased ATCA and the Parkes, Mopra, Hobart, Ceduna and Tidbinbilla telescopes. The maximum baseline of the LBA is $\sim1700\,$km, or five times smaller than that of the HSA. The sensitivity of a 12 hour observation is typically 20\,$\mu$Jy. The Hartebeestok telescope in South Africa can also be included to achieve an angular resolution comparable to the HSA but we do not consider it in this work due to the lack of intermediate baseline lengths. In determining the effective range of VLBI observations we consider three scenarios, outlined in Table \ref{tab:VLBI_properties};
\begin{enumerate}
    \item LBA observations of a source at $-30\degree$ declination;
    \item HSA observations of a source at $+20\degree$ declination and;
    \item Observations of a source at $+50\degree$ declination using the HSA without Arecibo
\end{enumerate}

\begin{figure*}
\includegraphics{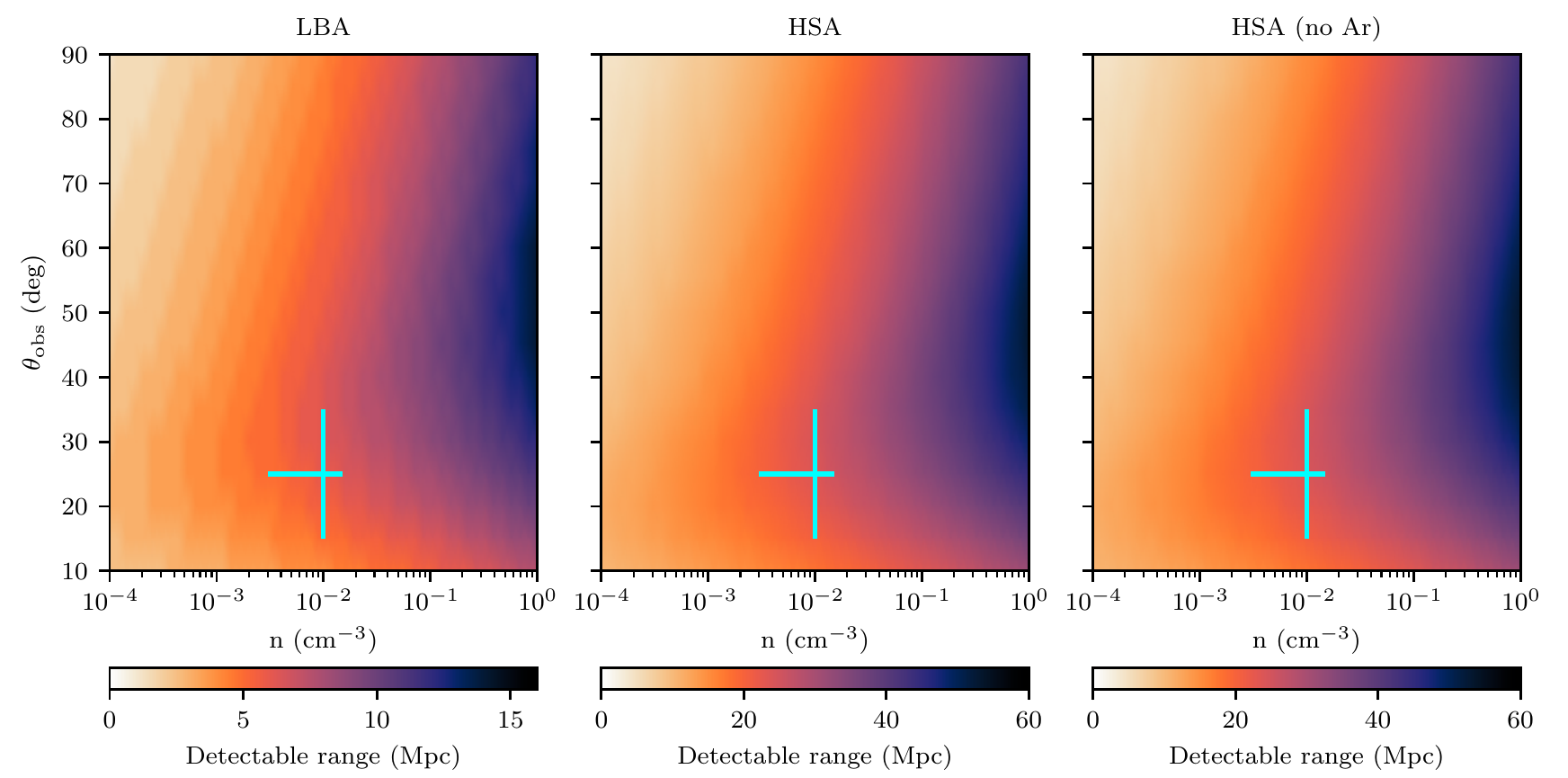}
\caption{Maximum distance at which outflow structure can be resolved for a range of circum-merger densities and merger inclination angles. Left: observations at $-30\degree$ declination with the LBA. Middle: observations at +20$\degree$ declination with the full HSA. Right: observations at +50$\degree$ declination using the HSA without Arecibo. \changes{The blue cross corresponds to the typical short GRB circum-merger density and estimates for the inclination angle of GW170817.}}
\label{fig:size_detectability}
\end{figure*}

\subsection{Resolving outflow structure}
For an afterglow to be resolvable with VLBI, it must be both sufficiently bright and sufficiently large. For sources near the detection threshold, only sizes comparable to or larger than  the VLBI synthesized beam can be measured with any degree of confidence, while brighter sources can be resolved even when smaller than the VLBI synthesized beam. The size of the source in the VLBI image is given by adding the source size, $\theta_{\rm S}$, and the VLBI beam size, $\theta_{\rm B}$, in quadrature;
\begin{equation}
    \Theta = \sqrt{\theta_{\rm B}^2 + \theta_{\rm S}^2}.
\end{equation}
The uncertainty in the fit of $\Theta$ given by
\begin{equation}
    \sigma_\Theta = \frac{\sqrt{2}\Theta}{\rho}
\end{equation}
where $\rho$ is the signal-to-noise ratio of the source given by a simplified form of Equation 41 from \citet{1997PASP..109..166C};
\begin{equation}
    \label{eq:rho}
    \rho^2 = \frac{\Theta^2}{4\theta_{\rm B}^2} \left[1+\left(\frac{\theta_{\rm B}}{\Theta}\right)^2\right]^3 \frac{S^2}{\sigma^2}
\end{equation}
where we have assumed the beam is a circular Gaussian. Equation \eqref{eq:rho} assumes that the signal-to-noise ratio is dominated by the thermal noise. However, the sensitivity of observations of bright sources are dominated by phase errors which limit the dynamic range of the image. We therefore impose an additional cutoff of $S/\sigma < 100$, corresponding to a phase error of $\sim 5\degree$ \citep{1999ASPC..180..275P}. The uncertainty in the source size can then be found using the equations of \citet{2017ApJ...839...35M};
\begin{equation}
    \sigma_{\theta_{\rm S}} = \sigma_{\Theta}\left[1 - \left(\frac{\theta_{\rm B}}{\Theta}\right)^2
    \right]^{-1/2}.
\end{equation}
For \changes{the source to be resolved and its angular size measured} we require that $\theta_{\rm S} > 2\sigma_{\theta_{\rm S}}$ and $\rho > 5$, and that both of these criteria are true for a minimum of 30 days.

In general the ability to resolve objects with large angular diameters may be limited by the structure and surface brightness of the object. However, we find that most events that are resolvable will only be slightly larger than the size of the beam, and therefore more detailed surface brightness considerations are not required.

Figure \ref{fig:size_detectability} shows the maximum distance at which events are resolvable as a function of circum-merger density and inclination angle. The detectability range is strongly dependent on merger parameters, and typical values are 20\,Mpc and 5\,Mpc using the HSA and LBA respectively \changes{(maximum detectable range 52\,Mpc and 14\,Mpc). Therefore only the very closest events will have resolvable outflows}. We also note that while events with low circum-merger densities will expand to have larger physical sizes than those with high circum-merger densities, and the corresponding source luminosity is much lower. We find that events occuring in denser environments will be more easily resolvable, as the dominant factor is the signal-to-noise of the source.

\changes{As expected, the merger energetics are the most dominant parameter, with the median range varying by a factor of 0.2--1.5 compared to the range for the fiducual parameters. Dense, on-axis events are the least affected by varying merger energetics, varying by a factor of 0.3--1.2, compared to low density off-axis events that vary by a factor of 0.08--1.9. Microphysics parameters are generally less dominant, varying by factors of 0.5--1.6 and 0.4--1.2 for $\epsilon_e$ and $\epsilon_B$ respectively.}

\subsection{Astrometric accuracy}
Our ability to measure centroid motion is strongly affected by the precision with which we can measure the position of the source in each epoch. There is no analytic solution to the astrometric accuracy of VLBI observations, and we are therefore limited to estimates using numerical simulations. \citet{2006A&A...452.1099P} provide values of systematic astrometric accuracy of VLBA and EVN observations for a range of declinations from $-25\degree$ to $+85\degree$ taking into account uncertainty in the Earth's orientation, calibrator position\footnote{Systematic errors in calibrator position will affect both epochs equally and can be ignored for our purposes.} and antenna positions, as well as uncertainties induced by the troposphere. For GW170817, \citet{2018Natur.561..355M} estimate the systematic contribution to astrometric uncertainty, $\theta_{\rm sys}$, with the HSA to be 0.15\,mas and 0.5\,mas in RA and Dec. respectively after taking into consideration ionospheric effects. However, these values are dominated by the low elevation of the source from the VLBA. We use more optimistic values, outlined for each of the three cases in Table \ref{tab:VLBI_properties}.

The astrometric accuracy also has a statistical component given by
\begin{equation}
    \label{eq:astrometric_stat}
    \theta_{\rm stat} = \frac{\theta_{\rm B}}{\sqrt{8\ln 2}\rho}
\end{equation}
and the total astrometric uncertainty is then
\begin{equation}
    \theta_{\rm total} = \sqrt{\theta_{\rm sys}^2+\theta_{\rm stat}^2}.
\end{equation}

\subsection{Detectability of centroid motion}
\label{subsec:centroid_detectability}
The intrinsic factors influencing the detectability of centroid motion are the merger luminosity and the magnitude of the observed source offset. A denser circum-merger medium will produce a more luminous afterglow that moves slower, while it will be harder to detect centroid motion in more distant events which have lower flux densities and smaller angular offsets. On-axis events have larger centroid offsets at early times when the outflow is relativistic as the apparent velocity is dominated by superluminal motion. However, as the outflow decelerates the apparent velocity becomes dominated by the transverse component of the physical velocity of the outflow, which is higher for off-axis events.

Figure \ref{fig:motion+flux} shows the offset of the afterglow centroid from the merger location and its flux density as a function of time for an event at a distance of 40\,Mpc with a circum-merger density of $n=10^{-3}\,$cm$^{-3}$, comparable to the parameters of GW170817 \changes{(see Figure \ref{fig:lit_comparison})}, for a range of inclination angles. The maximum detectable offset is the offset between the centroid position of the afterglow at the first and last times the flux density of the afterglow is above the detection threshold of the telescope.

\begin{figure*}
\includegraphics{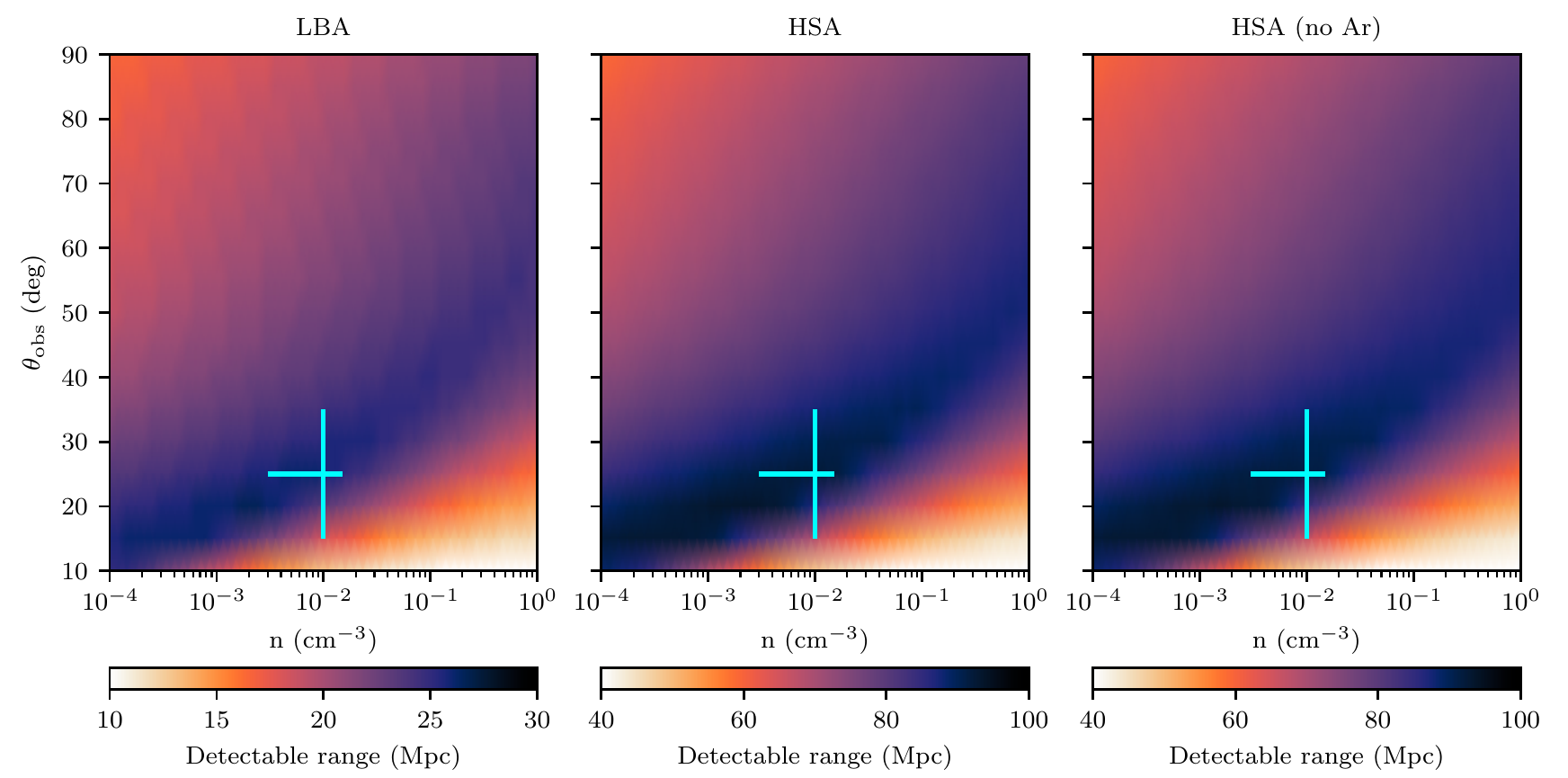}
\caption{Maximum distance at which centroid motion can be detected for a range of circum-merger densities and merger inclination angles. Left: observations at -30$\degree$ declination with the LBA. Middle: observations at +20$\degree$ declination with the full HSA. Right: observations at +50$\degree$ declination using the HSA without Arecibo. \changes{The blue cross corresponds to the typical short GRB circum-merger density and estimates for the inclination angle of GW170817.}}
\label{fig:offset_detectability_distance}
\end{figure*}

We calculate the detectable distance, $D$, by scaling the flux density of the afterglow by $D^{-2}$ and the angular offset by $D^{-1}$. We then calculate the centroid offset between each combination of times, $t_1$ and $t_2$, given by
\begin{equation}
    \langle X\rangle = |\langle x\rangle(t_2)-\langle x\rangle(t_1)|
\end{equation}
and define that offset as detectable if 
\begin{equation}
    \langle X\rangle > 5\left[\theta_{\rm total}(t_1)+\theta_{\rm total}(t_2)\right]
\end{equation}
Finally, for an event to be considered detectable we additionally impose the constraint that there must be at least a 30 day buffer around both $t_1$ and $t_2$, as observers will have minimal a priori knowledge of the optimal times to observe.

Figure \ref{fig:offset_detectability_distance} shows the detectable range as a function of circum-merger density and inclination angle. We find that the detectability of centroid motion is less dependent on merger parameters than the resolvability of the outflow. Typical ranges are 20\,Mpc and 80\,Mpc for the LBA and VLBA respectively, only 20\% lower than the maximum range. Unlike our ability to resolve outflow structure, the most dense and on-axis events have the lowest detectability ranges. While these events have the highest peak luminosities, the decline of the lightcurve is only weakly dependent on either parameter (e.g. see Figure \ref{fig:motion+flux}). Denser events have lower initial velocity, and also decelerate faster meaning that the apparent velocity of events becomes dominated by the physical velocity of the outflow (which is higher for off-axis events) at earlier times.

\changes{Again, merger energetics dominate the microphysics parameters, with the median range varying by a factor of 0.1--1.3. Regions with larger ranges saw a greater decrease for lower values of $E_{\rm iso}$, and a smaller increase for higher values of $E_{\rm iso}$. The median range varied by a factor of 0.5--1.3 for $\epsilon_e$ and 0.4--1.1 for $\epsilon_B$.}

Superluminal motion will not be detectable for a large fraction of events within the LIGO O3 detector range of 135\,Mpc \citep{2018LRR....21....3A}. As current gravitational wave detectors reach design sensitivity, a decreasing fraction of the newly accessible (more distant) events will exhibit centroid motion that is detectable by current VLBI facilities.

We note that the limiting factor of VLBI observations in general is the maximum baseline, which is constrained by the size of the Earth. While larger baselines can be achieved using space-VLBI satellites like \emph{RadioAstron} these facilities do not have sufficient sensitivity for gravitational wave follow-up \citep{2013ARep...57..153K}. Adding new telescopes to existing VLBI arrays and/or increasing observing bandwidth would improve sensitivity, increasing the detection range of events with low circum-merger densities and large inclination angles at current observing frequencies. The higher sensitivity could also be used to facilitate observations at higher frequency, making it possible to discern the smaller motion of more distant events (which would have otherwise been too faint at higher frequency). However, for many events, VLBI information will not provide significant additional constraints. We discuss the implications for the prospects for determining the value of H0 using VLBI techniques in Section \ref{subsec:H0}.

\section{Implications}
\subsection{Determining source size and geometry}
We have discussed two possible methods for determining the physical size of outflow from neutron star mergers. Observations of scintillation from afterglows will allow us to place constraints on a larger fraction of events than direct imaging with VLBI due to the detectable range being a factor of 2--3 larger. However we caution that both techniques have a role to play in understanding outflow geometry.

Scintillation observations enable us to place constraints on source size at early times and are more effective for dense, on-axis events, as the source remains compact for longer. In contrast, VLBI observations are more useful at later times when the outflow has expanded and the emission centroid has shifted away from the merger location. This distinction will be important for follow-up of events in the future. The late-time VLBI observations of GW170817 constrained the geometry of the merger outflow once it became jet-dominated \citep[tens of days post-merger;][]{2018Natur.561..355M,2019Sci...363..968G} but the early-time behaviour (and therefore the properties of the cocoon ejecta) is still not well understood.

As well as giving insight into different stages of source evolution, both of these techniques probe a different part of the parameter space. As the circum-merger density increases scintillation becomes more detectable, while either VLBI phenomena becomes less detectable. Comparing Figure \ref{fig:scintillation_distance} and \ref{fig:size_detectability} shows that while the detectability range for direct imaging of the outflow is only a few tens of megaparsecs, these observations can constrain source sizes for a significant fraction of the parameter space that is inaccessible for scintillation observations with current radio facilities.

Any constraints on source size will be useful in determining the structure of the merger outflow - e.g. \citet{2017Sci...358.1579H} used the absence of scintillation-induced variability to rule out the observed radio emission from GW170817 being produced by subrelativisitic ejecta. By obtaining a larger sample of neutron star mergers with well-understood outflow structure we may be able to shed light on the central engine that drives the outflow and produces short GRBs. Many models propose ways in which this central engine may be formed \citep[e.g. see][and references therein]{2019EPJA...55..132F}, but the exact mechanism is still unknown.

\subsection{Observations of superluminal motion} 
The measurement of centroid motion with VLBI provides strong evidence for the presence of a successful jet, and is a robust and direct method to determine the velocity of that jet (see also \citealt{linial2019} for an alternative method). Combining the observations of superluminal motion and afterglow light curve data, one can determine the jet half-opening angle and viewing angle \citep{2018Natur.561..355M,2019NatAs...3..940H}. If the lightcurve data only samples a single segment of the synchrotron spectrum \citep{1998ApJ...497L..17S}, the energetics of a jet still degenerates with the circum-merger density. However, if the synchrotron cooling break and/or self-absorption frequency are measured, this degeneracy can be broken. These parameters may be measured via multi-wavelength observations spanning radio to X-rays, although this was not possible in the follow-up of GW170817 because the synchrotron cooling break remained above the observable X-ray band \citep{2018ApJ...863L..18A}. Therefore, VLBI observations of neutron star mergers in conjunction with multi-wavelength lightcurve monitoring can potentially allow us to robustly determine the jet's energetics and structure, viewing angle, as well as the density of the surrounding medium.

In addition to determining merger parameters, centroid motion observations may potentially be used to infer the merger progenitors. For example, the mass ejection (and therefore electromagnetic emission) from neutron star-black hole mergers may by highly anisotropic in comparison to NS mergers which are expected to be axisymmetric \citep{kyutoku2013,2014PhRvD..90b4026F,2016PhRvD..94l3016F}.

While the link between neutron star mergers and short GRBs has been clearly demonstrated by the detection of GW170817 \citep{2017ApJ...848L..13A,2017ApJ...848L..14G}, the details of the relationship remains unclear. The inferred on-axis lightcurve of GW170817 is consistent with the known population of short GRBs occuring at cosmological distances \citep{2019ApJ...880L..23W,2019A&A...628A..18S}, suggesting that all short GRBs may be produced by the same mechanism and the observed diversity in the short GRB population is caused by extrinsic properties including the viewing angle \citep{2001ARep...45..236L,2002MNRAS.332..945R,2015ApJ...815..102F,2019arXiv190707599S}. Direct measurement of viewing angle and outflow energetics of a larger sample of events using VLBI observations will allow this claim to be tested and place tight constraints on the short GRB luminosity function.

VLBI observations of centroid motion can also be combined with radio lightcurve monitoring to infer the opening angle of the jet produced by the merger, as was done for GW170817 \citep{2018Natur.561..355M}. A sample of events with measured jet opening angles will constrain the inverse beaming
fraction of the GRBs, and thereby establish whether neutron star mergers are responsible for the entire short GRB population. Understanding the typical jet opening angle will also improve estimates of the rate of joint GRB-GW detections \citep{2019MNRAS.485.1435H,2019MNRAS.483..840B}, and inform future multi-messenger observing strategies.

\subsection{Hubble constant}
\label{subsec:H0}
We will be able to detect centroid motion in most events accessible with the HSA within 80\,Mpc, while the effective range for the rest of the sky is 20\,Mpc. Using the inferred neutron star merger rate (assuming a Gaussian mass distribution) from the first two LIGO observing runs \changes{\citep[$1090_{-800}^{+1720}$\,Gpc$^{-3}$yr$^{-1}$;][]{2020arXiv200101761T}}, we find that in future observing runs 1--2 events per year will be useful for combined radio-GW measurements of $H_0$. A second event would reduce the uncertainty in the measurement from \citet{2019NatAs...3..940H} to $\sim 5\%$, but resolving the tension between distance ladder and CMB measurements requires a precision of $<2\%$ which will not be achievable for decades.

In comparison, achieving this precision with only gravitational wave data from a localised event requires $\sim 100$ more events based on the precision achieved for GW170817 \citep{2017Natur.551...85A}. \changes{Including lightcurve modelling in this analysis can yield a improved precision \citep[e.g.][]{2017ApJ...851L..36G,2019arXiv191212218D,2020ApJ...888...67D}, although the total number of events required to resolve the $H_0$ tension is still $\sim 100$.}

Using the same inferred merger rate we expect to detect $\sim 25$ neutron star mergers per year during the fourth LIGO observing run based on a detector horizon of 190\,Mpc, and \changes{$\sim 125$} neutron star mergers per year in subsequent runs based on a detector horizon of 330\,Mpc \citep{2018LRR....21....3A}. If every merger is localised to a host galaxy then a measurement with sufficient precision to potentially resolve the $H_0$ tension will be achieved within the next decade. However as we have found LIGO/Virgo O3a\footnote{O3a is the first part of the third LIGO/Virgo observing run, from April-October 2019} it is unlikely that every detected merger will be localised as easily as GW170817. We also note that the signal-to-noise of a merger detection is strongly dependent on merger inclination (see \eqref{eq:gw_inc_angle}). Higher significance detections will have smaller localisation volumes and therefore face-on mergers occuring within the nominal 80\,Mpc VLBI range will be more likely to be localised to a host galaxy. However, as more gravitational wave detectors come online \citep{2018LRR....21....3A}, inclination angle measurements from gravitational wave data alone will improve and VLBI observations may become less useful. We also caution that measurement of $H_0$ using a population of events with VLBI-constrained inclination angles requires careful consideration of selection biases \citep{2019PhRvD.100j3523M}, which are not yet well understood, but will be as our sample of EM-bright gravitational wave events grows.

The uncertainty in the peculiar motion of the host galaxy of GW170817 is one of the largest errors in the current combined radio-GW  measurement of $H_0$ \citep{2019NatAs...3..940H,2019arXiv190908627M,2020MNRAS.492.3803H}. However, this uncertainty can be significantly reduced if similar measurements are done for merger events at farther distances. As shown in Figure \ref{fig:offset_detectability_distance}, the centroid motion can be measured by VLBI up to $\sim 100\,{\rm Mpc}$ in the case of the favorable density and viewing angle. Thus, detecting the centroid motion of the jet in such GW events will be particularly  important for combined radio-GW  measurements of $H_0$.

Another way to measure $H_0$, first proposed by \citet{1986Natur.323..310S} is the `dark siren' method where BBH localisation volumes are convolved with galaxy catalogues to get a probabilistic measurement of $H_0$ \citep{2012PhRvD..86d3011D}. This method was applied to GW170817 ignoring the knowledge of the host galaxy \citep{2019ApJ...871L..13F}, and achieves a similar result to \citet{2017Natur.551...85A}. \citet{2019ApJ...876L...7S} also apply this method to GW170814, and find $H_0 = 75_{-32}^{+40}$\,km\,$s^{-1}$\,Mpc$^{-1}$. \citet{2018Natur.562..545C} find that this method will only achieve a precision of $\sim 10\%$ within the next decade, while \citet{2018PhRvD..98b3502N} find that third generation GW detectors may achieve a precision of 7\% with only 25 BBH mergers. However, it will be possible to measure the redshift of BBH merger with third generation GW detectors to as good as 8\% \citep{2012PhRvL.108i1101M}, which will allow direct measurement of $H_0$ from gravitational wave events alone. The degeneracy between redshift and merger chirp mass may also be overcome with a large population of mergers and observational constraints on mass distributions \citep{2012PhRvD..85b3535T,2019ApJ...883L..42F}.

While the non-VLBI methods discussed above have larger uncertainties on a per-event basis, the larger number of available events reduces the uncertainty contribution from host galaxy peculiar velocities, which should be randomly oriented and therefore partially cancel each other out \citep{2020MNRAS.492.3803H}. It will also be possible to combine different standard siren measurements together, and while VLBI measurements may only be possible for $\sim 10\%$ of localised mergers, they will contribute more than that to the sensitivity of the overall measurement.

In general, we caution that these estimates of detection rates have large uncertainties due to small number statistics, the large uncertainty in the neutron star merger rate, and the even larger uncertainty in the distribution of circum-merger densities. They also rely on the assumptions of our detectability outlined in Section \ref{subsec:centroid_detectability} and the afterglow models from Section \ref{sec:geometry_models}.

\section{Conclusions}
In this paper we have discussed prospects for constraining the properties of neutron star mergers through observations of scintillation-induced variability and by using high resolution VLBI measurements to both detect motion of the emission centroid and directly image outflow structure. We find that while VLBI observations provide more direct measurements of source properties they are only feasible for the very closest events, while the scintillation technique can be applied to most events detected with current GW detectors it only provides indirect constraints on source size. Both techniques probe different parts of the merger parameter space, with VLBI measurements suited to events occuring in less dense environments. Additionally, both techniques probe different timescales and therefore where possible should be used in conjunction with one another to completely understand the source structure as the afterglow evolves. We also discuss prospects for measuring $H_0$ and resolving the tension between current competing measurements and find that while gravitational waves provide a completely independent technique that does not rely on distance ladders or complex statistical inferences, it will likely take at least a decade to achieve a precision that is comparable with current techniques. This improvement in precision relies not only on observing a larger population of mergers with VLBI-constrained inclination angles, but improvements in both hydrodynamic jet models and gravitational wave detector calibration.

\section*{Acknowledgements}
We thank Mark Walker for useful discussions\changes{, and both the anonymous referee and Gavin Lamb for a number of suggestions that improved this work}. DD is supported by an Australian Government Research Training Program Scholarship. TM acknowledges the support of the Australian Research Council through grant DP190100561. DLK was supported by NSF grant AST-1816492. Parts of this research were conducted by the Australian Research Council Centre of Excellence for Gravitational Wave Discovery (OzGrav), project number CE170100004. This research has made use of NASA's Astrophysics Data System Bibliographic Services.

\textit{Software:} Astropy \citep{2018AJ....156..123A}, Cmasher \citep{cmasher}, Jupyter \citep{Kluyver:2016aa}, Matplotlib \citep{2007CSE.....9...90H}, NE2001 \citep{2002astro.ph..7156C}, Numpy \citep{2011CSE....13b..22V}

%\bibliographystyle{mnras}
%\bibliography{bibliography}
\input{main.bbl}

\bsp
\label{lastpage}
\end{document}

%% file: lit_table.tex
\begingroup
\renewcommand{\arraystretch}{1.2}

\begin{table*}
    \caption{Estimates of the observing angle, $\theta_{\rm obs}$, jet opening angle, $\theta_j$, and circum-merger density, $n_0=n/({\rm 1\,cm}^{-3})$, microphysics parameters, $\epsilon_e, \epsilon_B$ and isotropic equivalent energy, $E_{{\rm iso},0}=E_{\rm iso}/({\rm erg})$, of GW170817 using Gaussian jet (GJ), boosted fireball (BF), power-law jet (PLJ) and other structured jet (SJ) models. We also include the time post-merger of the latest observation covered by the fit. We have calculated the estimate of $n$ from \citet{2019NatAs...3..940H} assuming an isotropic equivalent energy of $E_{\rm iso}=10^{52}$\,erg.}
    \label{tab:lit_comparison}
    \centering
    \begin{threeparttable}
    \begin{tabular*}{0.95\textwidth}{l @{\extracolsep{\fill}} cd{2.4}d{2.5}d{2.5}d{2.6}d{2.5}d{2.4}c}
        \hline\hline
           Reference &        Model &      \dcolhead{$\theta_{\rm obs}$} &            \dcolhead{$\theta_{j}$} &               \dcolhead{$\log_{10} n_0$} &           \dcolhead{$\log_{10} \epsilon_e$} &        \dcolhead{$\log_{10} \epsilon_B$} &       \dcolhead{$\log_{10} E_{{\rm iso},0}$} & Last obs.\\
       ~ & ~ &  \dcolhead{(deg)} & \dcolhead{(deg)} & ~ & ~ & ~ & ~ & (days)\\
       \hline
        \citet{2019Sci...363..968G}\tnote{*} &    GJ &    15.0^{+1.5}_{-1.0} &     3.4^{+1.0}_{-1.0} &     -3.6^{+0.7}_{-0.7} &              \dcolhead{-} &     -3.9^{+1.7}_{-1.5} &     52.4^{+0.6}_{-0.7} &                289 \\
            \citet{2019ApJ...886L..17H}\tnote{$\dagger$} &    BF &    30.4^{+4.0}_{-3.4} &     5.9^{+1.0}_{-0.7} &  -2.61^{+0.42}_{-0.63} &     -0.75^{+0.43}_{-0.62} &   -2.63^{+0.89}_{-1.2} &   52.33^{+0.6}_{-0.55} &                743 \\
 \citet{2019NatAs...3..940H}\tnote{*$\dagger$$\ddagger$} &   PLJ &  16.62^{+1.1}_{-0.57} &  3.44^{+0.57}_{-0.57} &  -4.03^{+0.17}_{-0.19} &              \dcolhead{-} &           \dcolhead{-} &           \dcolhead{-} &                294 \\
                                                       ~ &    GJ &  17.19^{+1.1}_{-0.57} &  2.75^{+0.17}_{-0.17} &   -4.06^{+0.19}_{-0.2} &              \dcolhead{-} &           \dcolhead{-} &           \dcolhead{-} &                294 \\
                             \citet{2019ApJ...870L..15L} &    SJ &    20.6^{+1.7}_{-1.7} &  4.01^{+0.57}_{-0.57} &     -3.3^{+0.6}_{-1.0} &        -1.3^{+0.6}_{-0.7} &     -2.4^{+1.4}_{-0.9} &     52.0^{+0.6}_{-0.9} &                358 \\
                                                       ~ &    GJ &    19.5^{+1.1}_{-1.1} &  5.16^{+0.57}_{-0.57} &     -4.1^{+0.5}_{-0.5} &        -1.4^{+0.5}_{-0.6} &     -2.1^{+0.8}_{-1.0} &     52.4^{+0.4}_{-0.5} &                358 \\
                             \citet{2018PhRvL.120x1103L} &    SJ &    33.0^{+4.0}_{-2.5} &   \dcolhead{$\sim 5$} &  -2.38^{+0.48}_{-0.21} &  -1.222^{+0.067}_{-0.079} &   -2.48^{+0.21}_{-0.4} &           \dcolhead{-} &                198 \\
                             \citet{2019MNRAS.485.2155L} &    GJ &    25.2^{+8.0}_{-5.7} &     4.6^{+1.7}_{-1.1} &     -2.5^{+1.1}_{-1.1} &      -1.28^{+0.81}_{-1.2} &     -4.1^{+1.4}_{-1.2} &   52.38^{+0.93}_{-0.9} &                360 \\
                             \citet{2018ApJ...867...57R} &    GJ &    26.9^{+8.6}_{-4.6} &     6.9^{+2.3}_{-1.7} &   -2.68^{+0.88}_{-1.0} &      -4.37^{+1.1}_{-0.48} &  -0.66^{+0.13}_{-0.45} &  51.76^{+0.52}_{-0.39} &                152 \\
                             \citet{2019arXiv190911691R} &    GJ &    22.9^{+6.3}_{-6.3} &     4.0^{+1.1}_{-1.1} &   -2.70^{+0.95}_{-1.0} &        -1.4^{+0.7}_{-1.1} &   -3.96^{+1.1}_{-0.74} &           \dcolhead{-} &                391 \\
                                                       ~ &   PLJ &    25.2^{+6.9}_{-7.4} &  2.86^{+0.57}_{-0.57} &     -2.6^{+1.1}_{-1.1} &      -1.24^{+0.73}_{-1.2} &   -3.76^{+1.1}_{-0.87} &           \dcolhead{-} &                391 \\
                             \citet{2019MNRAS.489.1919T} &    GJ &        29.^{+11}_{-12} &     4.6^{+1.7}_{-2.3} &   -2.37^{+0.84}_{-1.3} &     -1.13^{+0.53}_{-0.88} &  -4.18^{+0.85}_{-0.58} &           \dcolhead{-} &                391 \\
                             \citet{2019ApJ...880L..23W} &    BF &    30.3^{+7.0}_{-4.0} &   \dcolhead{$\sim 5$} &     -2.0^{+0.7}_{-1.0} &        -1.0^{+0.6}_{-0.9} &     -3.6^{+1.3}_{-1.4} &           \dcolhead{-} &                260 \\
        \hline\hline
    \end{tabular*}
    \begin{tablenotes}\footnotesize
    \item[*] Incorporates centroid motion measurements
    \item[$\dagger$] Does not incorporate optical data
    \item[$\ddagger$] Does not incorporate X-ray data
    \end{tablenotes}
    \end{threeparttable}
\end{table*}
\endgroup